\documentclass[prd]{revtex4}

\usepackage{amsmath}
\usepackage{bm}
\usepackage{graphics}
\usepackage{color}

\def\gsim{\;\rlap{\lower 2.5pt
 \hbox{$\sim$}}\raise 1.5pt\hbox{$>$}\;}
\def\lsim{\;\rlap{\lower 2.5pt

   \hbox{$\sim$}}\raise 1.5pt\hbox{$<$}\;}

\def\be{\begin{equation}}
\def\ee{\end{equation}}
\def\ba{\begin{eqnarray}}
\def\ea{\end{eqnarray}}

\begin{document}

\title{ Revisit relic gravitational waves   based on the latest CMB observations  }

\author{Minglei Tong}

\thanks{Email: mltong@ntsc.ac.cn}
\affiliation{ Korea Astronomy and Space Science Institute, Daejeon, 305-348, Korea}
\affiliation{Key Laboratory of Time and Frequency Primary Standards, National Time Service Center, Chinese Academy of Sciences, Xi'an, Shaanxi 710600,  China }

\begin{abstract}
 According to the CMB observations, Mielczarek (\cite{Mielczarek}) evaluated the reheating temperature,
 which could help to determine the history of the Universe. In this paper, we recalculate the
 reheating temperature using the new data from WMAP 7 observations. Based on that, we list the approximate solutions
 of relic gravitational waves (RGWs) for various frequency bands.
 With the combination of the quantum normalization of RGWs when they are produced and
 the CMB observations, we obtain the relation  between the tensor-to-scalar ratio $r$ and the inflation index $\beta$
 for a given scalar spectral index $n_s$. As a comparison, the diagram $r-\beta$ in the slow-roll inflation model
 is also given. Thus, the observational limits of $r$ from CMB lead  to the constraints on the
 value of $\beta$. Then, we illustrate  the energy density spectrum of RGWs
 with the quantum normalization for different values of $r$ and the corresponding $\beta$. For comparison,
 the energy density spectra of RGWs with parameters based on slow-roll inflation are also discussed. We find that
 the values of $n_s$ affect the spectra of RGWs sensitively in the very high frequencies. Based on the current and
 planed gravitational wave detectors, we discuss the detectabilities of RGWs.

\

\end{abstract}

\maketitle

\large

\section{ Introduction}

General relativity and quantum mechanics  predict a stochastic background of relic
gravitational waves (RGWs)
\cite{grishchuk1,grishchuk,grishchuk3,starobinsky,Maggiore,Giovannini}, generated during the
early inflationary stage. The primordial  amplitudes could be determined by
the quantum normalization at the time of the wave modes crossing the horizon during the inflation.
 After that, the evolution of RGWs are mainly determined
by a sequence of stages of cosmic expansion
including the current acceleration \cite{zhang2,Zhang4},
 since the interaction of RGWs with other cosmic components
is typically very weak. Therefore, RGWs carry a unique information
of the early Universe, and serve as a probe into the Universe much
earlier than the recombination stage. As an interesting source for gravitational wave (GW) detectors,
RGWs exist everywhere and anytime  unlike GWs radiated by usual astrophysical process.
Moreover,
RGWs spread a very broad range of frequency, $10^{-18}-10^{10}$ Hz, making themselves become
  one of the major scientific goals of various GW detectors with different response frequency bands.
  The current and planed GW detectors contain  the ground-based interferometers,
such as  LIGO \cite{ ligo1}, Advanced LIGO \cite{ligo2,advligo},
VIRGO \cite{virgo,virgocurve}, GEO \cite{geo}, AIGO \cite{Degallaix}, LCGT \cite{lcgt} and ET \cite{Punturo,Hild} aiming at the frequency range $10^2-10^3$ Hz;
the space interferometers, such as the future LISA \cite{lisa0,lisa} which is sensitive in the frequency range $10^{-4}-10^{-1}$ Hz,  BBO \cite{Crowder,Cutler}  and DECIGO
 \cite{Kawamura} which  both are sensitive in the
frequency range $0.1-10$ Hz; and the pulsar timing array, such as  PPTA \cite{PPTA,Jenet} and the planned
SKA \cite{Kramer} with the frequency window $10^{-9}-10^{-6}$ Hz.
Besides, there some potential very-high-frequency GW detectors, such as
the waveguide detector \cite{cruise},
the proposed gaussian maser beam detector around GHz \cite{fangyu}, and the 100 MHz detector with a pair of
75-cm baseline synchronous recycling  interferometers \cite{Akutsu}.
Furthermore,   the very low frequency portion of
RGWs also contribute to the  anisotropies
and polarizations of cosmic microwave background (CMB) \cite{basko},
yielding a magnetic type polarization of CMB
as a distinguished signal of RGWs.
WMAP \cite{Peiris,Spergel,Hinshaw,Komatsu},
Planck \cite{Planck}, the ground-based  ACTPol  \cite{Niemack}  and the proposed CMBpol \cite{CMBpol}
are of this type.

The reheating temperature, $T_{\rm{RH}}$, carries rich information of the early Universe, and   relates
to the decay rate of the inflation as $T_{\rm{RH}}\propto \sqrt{\Gamma}$ \cite{Kolb,Nakayama}
Recently, the reheating temperature    was evaluated  \cite{Mielczarek} according to  the CMB observations by WMAP 7
in the frame of the slow-roll  inflation \cite{Komatsu}. Then, the
expansion histories  of different stages could be determined subsequently. In this paper, we reevaluate  the reheating temperature using the
latest observational data from CMB, and adopt the resulting expansion periods of different
phases of the Universe as references. The referenced reheating temperature can help us to
divide the phases of the Universe definitely.
The evolutions of the RGWs at various phases can   be determined subsequently, and the primordial  amplitude
was normalized due to the quantum condition during inflation \cite{grishchuk,grishchuk3}. For the
present time, the solutions of RGWs can be obtained for  different frequency bands corresponding to
the modes re-entered the horizon at different phases. Note that, this model of RGWs is free from the slow-roll
inflation. Therefore, the above reheating temperature based on the slow-roll inflation just serves as a reference.
On the other hand, the anisotropies due to the tensor metric perturbations (gravitational waves) can be scaled
to those due to the observations of the scalar perturbations by introducing a parameter $r$ called tensor-to-scalar
ratio. Combining the observations of the CMB   and the quantum normalization of RGWs when they are generated,
a constraint condition is arrived between  the ratio $r$, the inflation index $\beta$, and the  index  $\beta_s$ describing the
expansion behavior of the Universe from the end of inflation
 to the reheating process. For the chaotic inflation with a quadratic potential $V=\frac{1}{2}m^2\phi^2$, which means  $\beta_s=1$,  the diagram $r-\beta$ will be illustrated for given values of the scalar spectral index $n_s$. The resulting  spectra of RGWs for
 different values of $r$ the corresponding $\beta$ will be demonstrated.
 For comparison, we will also discuss the spectra of RGWs with the values of $r$ and $\beta$ predicted by the
 slow-roll inflation itself.
To this end, the spectra of RGWs given by different models and different parameters will
confront  the various current and planed GW detectors.

The outline of this paper is as follows.
In Sec. II, we recalculate the reheating temperature
using the latest data from CMB and plot it as a function of the scalar
spectral index. Based on that, the scale factor $a(\tau)$ is specified
for consecutive stages of cosmic expansion.
In section III, we present the resulting approximate solutions of the
spectrum of RGWs for various frequency bands.
In section IV, the spectra of RGWs for different values of  parameters are
 shown and some comparisons between the spectra based on quantum normalization and those
 based on slow-roll inflation will be given.
 Some discussions are summarized in Sec. V.  Throughout this paper,  we use the units   $c=\hbar=k_B=1$.
Indices $\lambda$, $\mu$, $\nu$,... run from 0 to 3, and $i$, $j$, $k$,... run from 1 to 3.

\section{  The expansion history of the universe}

For a spatially flat ($k=0$)  universe
the Robertson-Walker spacetime has a metric
\be
ds^2=a^2(\tau)[-d\tau^2+\delta_{ij}dx^idx^j],
\ee
where $\tau$ is the conformal time, and the scale factor  $a(\tau)$
is described by the following successive stages
\cite{grishchuk,zhang2}:

The inflationary stage:
\be \label{inflation}
a(\tau)=l_0|\tau|^{1+\beta},\,\,\,\,-\infty<\tau\leq \tau_1,
\ee
where the inflation index $\beta$ is an important model parameter.
The special case of $\beta=-2$  corresponds the exact de Sitter expansion.
However, both the model-predicted and the observed results
indicate that the value of $\beta$ can differ slightly from  $-2$.

The preheating stage :
\be\label{preheat}
a(\tau)=a_z|\tau-\tau_p|^{1+\beta_s},\,\,\,\,\tau_1\leq \tau\leq \tau_s,
\ee
where the parameter  $\beta_s$ is usually taken as a constant. After inflation, the inflation
field undergoes coherent oscillations at the bottom of potential well. The reheating process
takes place the Hubble parameter $H$
falls to the value of the inflation decay rate $\Gamma_\phi$, if we assume that
the reheating  is instantaneous. Then the preheating process lasts from
the end of inflation to  the happening of reheating. Below, we will discuss the value of $\beta_s$
in detail.

The radiation-dominant stage :
\be \label{r}
a(\tau)=a_e(\tau-\tau_e),\,\,\,\,\tau_s\leq \tau\leq \tau_2.
\ee

The matter-dominant stage:
\be \label{m}
a(\tau)=a_m(\tau-\tau_m)^2,\,\,\,\,\tau_2 \leq \tau\leq \tau_E.
\ee

The accelerating stage up to the present time $\tau_0$
\cite{zhang2}:
\be \label{accel}
a(\tau)=l_H|\tau-\tau_a|^{-\gamma},\,\,\,\,\tau_E \leq \tau\leq
\tau_0,
 \ee
 where $\gamma$ is a $\Omega_\Lambda$-dependent
parameter, and $\Omega_\Lambda$ is the energy density contrast.
To be specific, we take $\gamma\simeq 1.97$
\cite{zhangtong} for $\Omega_{\Lambda}=0.73 $ \cite{Komatsu} in this paper.
It is convenient to choose the normalization  $|\tau_0-\tau_a|=1$, i.e.,
the present scale factor $a(\tau_0)=l_H$. From the definition of the
Hubble constant, one has $l_H=\gamma/H_0$,
where $H_0=100\, h$ km s$^{-1}$Mpc$^{-1}$ is the present Hubble constant.
We  take  $h\simeq 0.704$ \cite{Komatsu} throughout this paper.
Imposing $\beta$ and $\beta_s$ as  model parameters,
 only 12 constants remain in all the expressions of the scale factor.
The continuity of $a(\tau)$ and $a'(\tau)$ at the four given
joining points $\tau_1$, $\tau_s$, $\tau_2$ and $\tau_E$ provide 8 constraints.
If we also know the expansion history of various stages, i.e., the definite values of
$\zeta_1\equiv{a(\tau_s)}/{a(\tau_1)}$,
$\zeta_s\equiv{a(\tau_2)}/{a(\tau_s)}$,
$\zeta_2\equiv{a(\tau_E)}/{a(\tau_2)}$,
and
$\zeta_E\equiv{a(\tau_0)}/{a(\tau_E)}$, all the 12 constants can be fixed
completely \cite{Miao}.

\begin{figure}
\resizebox{100mm}{!}{\includegraphics{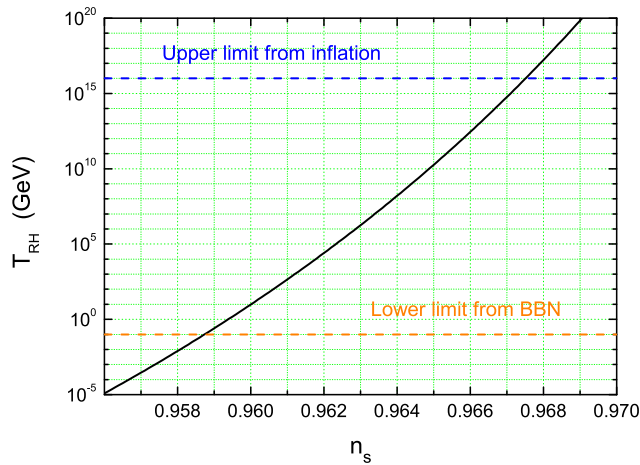}}
\caption{\label{reheat}
The reheating temperature $T_{\rm{RH}}$ as a function of the scalar spectral index $n_s$, where $\beta_s=1$ and $g_{\star s}=3.91$ are set.
}
\end{figure}

 The late-time universe is well know. For the $\Lambda$CDM model,
the  density  of dark energy,  which drives the accelerating expansion of the universe,
is constant. Based on that, one easily has
$\zeta_E =1+z_E=({\Omega_\Lambda}/{\Omega_m})^{1/3}\simeq1.4$, where $z_E$ is
the redshift when the  accelerating expansion begins.
 For the duration of the matter-dominated
stage, we get straightforwardly $\zeta_2
             =\frac{a(\tau_0)}{a(\tau_2)} \frac{a(\tau_E)}{a(\tau_0)}
=(1+z_{eq}) \zeta_E^{-1}$ with $z_{eq}=3240$ \cite{Komatsu}.
However, the histories of the radiation-dominated stage and the preheating stage are
not known well. Recently, Mielczarek \cite{Mielczarek} proposed a method to
evaluate the reheating temperature, $T_{\rm{RH}}$, under the frame of the slow-roll inflation model combing the observations
from WMAP. Using this method, one can easily obtain the following expression:
\be\label{reheating}
T_{\rm{RH}}=\frac{15\,m_{Pl}}{8\,g_{\star s}\,\pi^{7/2}}\sqrt{\frac{1-n_s}{A_s}}\left(\frac{k_0^p}{T_{\rm{CMB}}}
\right)^3\exp{\left(\frac{6}{1-n_s}\right)}
\ee
where $m_{Pl}\equiv 1/\sqrt{G}=1.22\cdot10^{19}$ GeV is the Plank mass, $g_{\star s}=3.91$ counts  the effective number of photons plus three species of massless neutrinos contributing to the radiation entropy  during the recombination \cite{Yuki},  $n_s$ is the scalar spectral index, $A_s=(2.43\pm 0.11)\cdot10^{-9}$  is the amplitude
of the scalar power spectrum at the pivot physical  wavenumber $k_0^p=0.002$ Mpc$^{-1}$ \cite{here}, and
$T_{\rm{CMB}}=2.725\ {\rm{K}}=2.348\cdot10^{-13}$ GeV is the present temperature of CMB.
 Note that, in deriving Eq. (\ref{reheating}), the approximation of $\beta\approx-2$ was used, and a quadratic potential $V(\phi)=\frac{1}{2}m^2\phi^2$ of the scalar field, where the scalar mass $m$ can be fixed to be $\sim 10^{13}$ GeV \cite{Kuroyanagi,Mielczarek} by CMB observations, was assumed. As analyzed by Turner \cite{Turner}, during the prehearting stage, when the oscillating frequency of $\phi$ is much greater than the expansion rate of the Universe,  the coherent scalar field oscillations behave like a fluid with $p=w\rho$, where the equation of state $w$ depends upon the form of the scalar potential $V(\phi)$. Say $V(\phi)=\lambda \phi^n$, one has $w=\frac{n-2}{n+2}$ and
$\rho$ decreases as $a^{-6n/(n+2)}$. For $V(\phi)=\frac{1}{2}m^2\phi^2$, i.e., $n=2$, one has $w=0$ and $\rho\propto a^{-3} $. This result was also verified by Martin and Ringeval \cite{martin} using numerical method, and it was found that the average $w$ never deviates from zero exceeding $8\%$. Hence, during the   evolution of the prehearting stage,  the energy
density  drops approximately in the same way as  the matter-dominated Universe \cite{Starobinsky2},
 and the scale factor evolves as $a(t)\propto t^{2/3}$,  which means $\beta_s=1$ from Eq. (\ref{preheat}). So, in the following we set $\beta_s=1$ except that we write it explicitly.
Moreover, the following relations in the slow-roll inflation \cite{LiddleA}:
\be\label{slow}
\epsilon\equiv\frac{m^2_{Pl}}{16\pi}\left(\frac{V'}{V}\right)^2, \qquad
\eta\equiv\frac{m^2_{Pl}}{8\pi}\left(\frac{V''}{V}\right),\qquad
n_s=1-6\epsilon+2\eta,
\ee
were also used in deriving Eq. (\ref{reheating}).
    According to  the observations
of WMAP 7, Mielczarek obtained $T_{\rm{RH}}=3.5\cdot10^6$ GeV  with  $n_s=0.963\pm0.012$ and $A_s=2.441^{+0.088}_{-0.092} \cdot 10^{-9}$, assuming $g_{\star s}=2$ being the effective number of relativistic degree of freedom only contributed
by photons during recombination.
One can recalculate $T_{\rm{RH}}$ using the updated data
given by WMAP 7 \cite{Komatsu}. For example, for WMAP Seven-year Mean, $n_s=0.967\pm0.014$, one
has  $ T_{\rm{RH}}\simeq5.8\cdot10^{14}\ {\rm GeV}$ with a relative uncertainty $\sigma(T_{\rm{RH}})/T_{\rm{RH}}\approx77$;
and for WMAP Seven-year ML, $n_s=0.966$, one has
$T_{\rm{RH}}\simeq2.8\cdot10^{12}\ {\rm GeV}$.
Thus, one can see that $T_{\rm{RH}}$ depends on the value of $n_s$ very sensitively. This is because the expression
of $T_{\rm{RH}}$ contains a exponential factor $\exp{(\frac{6}{1-n_s})}$, which is dependent on the value of $n_s$ very
  sensitively.
   For a more general potential with power-law form $V=\lambda\phi^n$, one can easily obtain the $e$-foldingor number $N_{\rm{obs}}\simeq\frac{n+2}{2(1-n_s)}$ using Eq. (14) in Ref. \cite{Mielczarek}, and in turn the resultant $T_{\rm{RH}}$ has an exponential factor $\exp{(\frac{3(n+2)}{2(1-n_s)})}$. For example, $n=4$, which implies $w=1/3$ and $\beta_s=0$,  $T_{\rm{RH}}$ contains a factor $\exp{(\frac{9}{1-n_s})}$. Therefore, $T_{\rm{RH}}$ also depends on the power index $n$, and in turn, on $\beta_s$ very sensitively. We plot the reheating temperature as a function of the scalar
  spectral index in Fig. \ref{reheat}.
From the bottom, the big bang nucleosynthesis (BBN) gives a  constraint of the reheating temperature
$T_{\rm{RH}}\gtrsim10$ MeV \cite{hannestad}. From the top, the constraint comes from the energy scale
at the end of inflation $T_{\rm{RH}}\lesssim10^{16}$ GeV. Under these constraints, one can see
that the scalar spectral index should locate at the range of $0.959\lesssim n_s\lesssim 0.967$.

 After the instantaneous reheating,
the universe is filled with the relativistic plasma. We assume the expansion of the relativistic gas to be
adiabatic, which is valid until the entropy transfer between the radiation and other components can be neglected.
Therefore,  the conservation of the entropy gives the increase of the scale factor from the reheating till  the recombination \cite{Mielczarek},
\be\label{delta1}
\frac{a_{rec}}{a(\tau_s)}=\frac{T_{\rm{RH}}}{T_{rec}}\left(\frac{g_{\ast s}}{g_{\star s}}\right)^{1/3},
\ee
where $a_{rec}$  and $T_{rec}$ stand  for the scale factor and the temperature at the recombination, respectively.
 $g_{\ast s}$ counts the effective number of relativistic species contributing to the entropy during the reheating. In the standard model of elementary particles,
 one has $g_{\ast}=g_{\ast s}=106.75$ at the energy scale above $\sim 1$ TeV \cite{Yuki}, where $g_{\ast}$ counts the effective number of relativistic species contributing to the energy density during the reheating.
 Moreover, as pointed in \cite{Mielczarek}, if the temperature of reheating is greater than the electroweak energy scale, $T_{\rm{RH}}\gtrsim300$ GeV, one may expect that $g_{\ast }\geq106.75$. Thus, in this paper we set $g_{\ast}=g_{\ast s}=106.75$.
 Based on Eq. (\ref{delta1}), one easily obtain
\be\label{zetas}
\zeta_s=\frac{a(\tau_2)}{a_{rec}}\frac{a_{rec}}{a(\tau_s)}=\frac{T_{\rm{RH}}}{T_{CMB}(1+z_{eq})}\left(\frac{g_{\ast s}}{g_{\star s}}\right)^{1/3},
\ee
where we have used $T_{rec}=T_{CMB}(1+z_{rec})$.
Finally, for the slow-roll inflation, the increase of the scale factor during the preheating stage with $\beta_s=1$ is given by \cite{Mielczarek}
\be\label{delta2}
\zeta_1=\left(\frac{15m^4_{Pl}A_s(1-n_s)^2}{64\pi^2g_{\ast }T_{\rm{RH}}^4}\right)^{1/3},
\ee
 Even though $\zeta_s$ and $\zeta_1$ are dependent on $g_{\ast}$, $T_{\rm{RH}}$ is independent on
 $g_{\ast}$ as shown in Eq. (\ref{reheating}). For WMAP Seven-year Mean, $n_s=0.967$, one
has  $\zeta_s\simeq2.3\cdot10^{24}$  and $\zeta_1\simeq4.86$; while for  WMAP Seven-year ML, $n_s=0.966$, one has
$\zeta_s\simeq1.11\cdot10^{22}$  and $\zeta_1\simeq6.07\cdot10^3$. In the following, the resulting $\zeta_s$ and $\zeta_1$ under the
frame of the slow-roll inflation will
serve as referenced tools to obtain the solutions of the RGWs, even though some calculations of RGWs based on
quantum normalization  are free from the
slow-roll inflation.

\section{   RGWs in the accelerating universe}

In the presence of
the gravitational waves,
the perturbed Robertson-Walker metric  is given by
\be
ds^2=a^2(\tau)[-d\tau^2+(\delta_{ij}+h_{ij})dx^idx^j],
\ee
where the tensorial perturbation $h_{ij}$
is traceless $h^i_{\,\,i}=0$ and transverse $h_{ij,j}=0$.
It can be decomposed into the Fourier $k$-modes and into the polarization
states, denoted by $\sigma$,  as
\be
\label{planwave}
h_{ij}(\tau,{\bf x})=
   \sum_{\sigma=+,\times}\int\frac{d^3k}{(2\pi)^{3/2}}
         \epsilon^{(\sigma)}_{ij}h_k^{(\sigma)}(\tau)e^{i\bf{k}\cdot{x}},
\ee
where
$h_{-k}^{(\sigma)*}(\tau)=h_k^{(\sigma)}(\tau)$
ensuring that $h_{ij}$ be real, the comoving wave number $k$ is related with
the wave vector $\mathbf{k}$ by $k=(\delta_{ij}k^ik^j)^{1/2}$, and the polarization
tensor  $\epsilon^{(\sigma)}_{ij}$ satisfies \cite{grishchuk}:
\be
\epsilon^{(\sigma)}_{ij}\epsilon^{(\sigma'){ij}}=2\delta_{\sigma\sigma'}, \quad
\epsilon^{(\sigma)}_{ij}\delta^{ij}=0, \quad
\epsilon^{(\sigma)}_{ij}n^j=0,\quad
\epsilon^{(\sigma)}_{ij}(-\mathbf{k})=\epsilon^{(\sigma)}_{ij}(\mathbf{k}).
\ee
In terms of the mode $h^{(\sigma)}_{k}$,
the wave equation is
\be \label{eq}
h^{ (\sigma) }_{k}{''}(\tau)
+2\frac{a'(\tau)}{a(\tau)}h^{ (\sigma) }_k {'}(\tau)
+k^2 h^{(\sigma)}_k(\tau )=0,
\ee
where a prime means taking derivative with respect to  $\tau$.
The two polarizations of
$h^{(\sigma)}_k(\tau )$ have the same statistical properties and give equal contributions to the unpolarized RGWs background,
so the super index $(\sigma)$ can be dropped.
Introducing a new notation \cite{grishchuk1,grishchuk,zhang2}: $\mu_k(\tau)\equiv a(\tau) h_k(\tau)$, Eq.(\ref{eq})  reduces to
\be\label{ueq}
\mu_k''+\left(k^2-\frac{a''}{a}\right)\mu_k=0.
\ee
For the high-frequency limit, i.e., the term $a''/a$ can be neglected, the solution to Eq.(\ref{ueq}) has the usual oscillatory from:  $\mu_k(\tau)=A_k e^{-ik\tau}+B_ke^{ik\tau}$, where the constants $A_k$ and $B_k$ are determined by the initial conditions,
and $h_k$ decreases  adiabatically with the expansion of the universe:
\be\label{inhorizon}
h_k\propto \frac{e^{\pm ik\tau}}{a(\tau)}.
\ee
 On the other hand,
when the term $a''/a$ is dominant, the dominant solution of Eq.(\ref{ueq}) is $\mu_k(\tau)=C_ka(\tau)$ for growing functions $a(\tau)$ in expanding universes, and
\be\label{outhorizon}
h_k=const.
\ee
In other words, when the wavelength of the mode $h_k(\tau)$,
 $\lambda={2\pi a}/{k}$, is much less than the horizon of the universe,
  $1/H=a^2(\tau)/a'(\tau)$, $h_k(\tau)$ decays with the expansion of the universe;
  while if $\lambda\gg 1/H$,  $h_k(\tau)$ will keep constant \cite{grishchuk,grishchuk1}.
   The wavelengths of various modes $h_k$ are stretched outside the horizon during the inflation stage,
   and they will be constant until re-enter the horizon again.
    When the mode $h_k$ re-enters the horizon, it will decay as $h_k\propto
    1/a(\tau)$.

The spectrum of RGWs $h(k,\tau)$ is defined by
\be \langle
h^{ij}(\tau,\mathbf{x})h_{ij}(\tau,\mathbf{x})\rangle\equiv\int_0^\infty
h^2(k,\tau)\frac{dk}{k},
\ee
where the angle brackets mean ensemble
average. The dimensionless spectrum $h(k,\tau)$ relates to the mode $h_k(\tau)$ as
\cite{TongZhang} \be \label{relation0}
h(k,\tau)=\frac{\sqrt{2}}{\pi}k^{3/2} |h_k(\tau)|.
\ee
The
primordial spectrum of RGWs at the time $\tau_i$ of the
horizon-crossing during the inflation has a power-law form
\cite{grishchuk,grishchuk3,zhang2,Miao}:
\be \label{initialspectrum}
h(k,\tau_i) =A\left(\frac{k}{k_H}\right)^{2+\beta},
\ee
where the constant $A$ representing the initial condition is to be determined both
from theories and observations.  An exact de Sitter expansion, i.e., $\beta=-2$,  yields   a  scale-invariant spectrum.
   Consider Eqs. (\ref{inhorizon}), (\ref{outhorizon}) and (\ref{relation0}), one knows that
the  spectrum  $h(k,\tau)$ will decay  as $\propto 1/a(\tau)$ when it reenter  the cosmic horizon, $1/H$.
In the following, let us discuss the properties of the present RGWs spectrum $h(k,\tau_0)$
with different frequency bands.
 The characteristic comoving wave number at a certain conformal time $\tau_x$ is give by
 \be\label{wavenumber}
k_x\equiv k(\tau_x) = \frac{2\pi a(\tau_x)}{1/H(\tau_x)}.
\ee
It is easily to obtain $k_H= 2\pi \gamma$. Explicitly, one has following relations:
\be\label{frelation}
  \frac{k_E}{k_H}
    = \zeta_E^{-\frac{1}{\gamma}},\quad
    \frac{k_2}{k_E}=\zeta_2^{\frac{1}{2}},\quad
    \frac{k_s}{k_2}=\zeta_s, \quad
    \frac{k_1}{k_s}=\zeta_1^{\frac{1}{1+\beta_s}}.
 \ee

\begin{figure}
\resizebox{100mm}{!}{\includegraphics{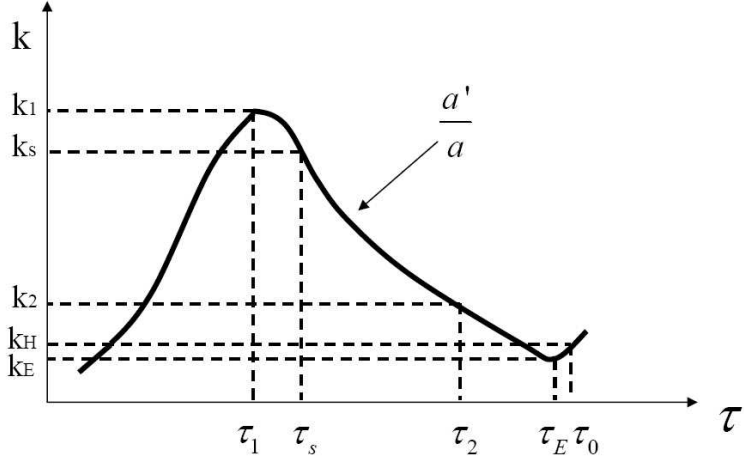}}
\caption{\label{represent}
The evolution of $a'/a.$
}
\end{figure}

As shown in Fig.\ref{represent}, for the comoving wave number $k<k_E$, the modes of RGWs
have been  outside the horizon all the time. Thus, these modes never decay and
keep their original amplitudes at present. For, $k_E<k<k_H$, the modes entered the horizon before
the  beginning of the acceleration and went out the horizon again before the present time.
We denote the time of the mode $k$ entering and going out of  the horizon  as $\tau_\ast$ and $\tau_{\ast\ast}$,
 respectively. The  corresponding scale factor are marked as $a_\ast$ and $a_{\ast\ast}$, respectively.
 Then, the present spectrum for $k_E<k<k_H$ can be written as
 \be
 h(k,\tau_0)=A\left(\frac{k}{k_H}\right)^{2+\beta}\frac{a_\ast}{a_{\ast\ast}}
 \ee
 According to Eqs.(\ref{m}), (\ref{accel}) and (\ref{wavenumber}),  one can easily obtain $a_\ast/a(\tau_E)=(k_E/k)^2$
 and $a(\tau_E)/a_{\ast\ast}=(k/k_E)^{-\gamma}$, which lead to $h(k,\tau_0)=h(k,\tau_i)(k/k_E)^{-(2+\gamma)}$.
 For $k>k_H$, after the modes entering the horizon, they have been inside the horizon up to now. Therefore, the spectrum
 for all the modes of $k>k_H$ has the following form:
 \be
 h(k,\tau_0)=A\left(\frac{k}{k_H}\right)^{2+\beta}\frac{a_\ast}{a(\tau_0)},
 \ee
 where we have still used the notes $a_\ast$ standing for the scale factor when the $k-$mode enters the horizon.
Using  the similar analysis given above, one can  get the present spectrum $h(k,\tau_0)$ for different bands of wave number which correspond to different stages of the universe.  Generally, we summarize the approximate solutions of $h(k,\tau_0)$ uniformly as follows ,
\ba\label{htaue}
&&h(k,\tau_0)=A\left(\frac{k}{k_H}\right)^{2+\beta}, \qquad k\leq k_E;\\
&&h(k,\tau_0)=A\left(\frac{k}{k_H}\right)^{\beta-\gamma}(1+z_E)^{-\frac{2+\gamma}{\gamma}}, \qquad  k_E\leq k\leq k_H;\\ \label{htauh}
&&h(k,\tau_0)=A\left(\frac{k}{k_H}\right)^{\beta}(1+z_E)^{-\frac{2+\gamma}{\gamma}}, \qquad  k_H\leq k\leq k_2;\\\label{hpulsar}
&&h(k,\tau_0)=A\left(\frac{k}{k_H}\right)^{1+\beta}\left(\frac{k_H}{k_2}\right)(1+z_E)^{-\frac{2+\gamma}{\gamma}}, \qquad  k_2\leq k\leq k_s;\\ \label{hf1}
&&h(k,\tau_0)=A\left(\frac{k}{k_H}\right)^{1+\beta-\beta_s}\left(\frac{k_s}{k_H}\right)^{\beta_s}\left(\frac{k_H}{k_2}\right)
(1+z_E)^{-\frac{2+\gamma}{\gamma}}, \qquad  k_s\leq k\leq k_1.
\ea
 The factor $(1+z_E)^{-\frac{2+\gamma}{\gamma}}$ means a reduction due to the accelerating expansion of the universe.
  The above equations reduce  to the corresponding results shown in \cite{zhang2},
  where  $\gamma=1$ was assumed.

  Another  important quantity often used in constraining
RGWs is its present energy density parameter defined by
$ \Omega_{GW}=<\rho_{g}>/{\rho_c}$,
where $\rho_g=\frac{1}{32\pi G}h_{ij,0}h^{ij}_{,0}$
is the energy density of RGWs,
and $\rho_c=3H_0^2/8\pi G$ is the critical energy density.
A direct calculation yields \cite{grishchuk3,Maggiore}
\be\label{gwe}
\Omega_{GW}=
\int_{f_{low}}^{f_{upper}} \Omega_{g}(f)\frac{df}{f},
\ee
with
\be\label{omega}
\Omega_{g}(f)=\frac{2\pi^2}{3}
        h^2_c(f)
     \Big(\frac{f}{H_0}\Big)^2
\ee
being the dimensionless  energy density spectrum. We have used  a new notation, $h_c(f)=h(f,\tau_0)/\sqrt{2}$,
called {\it characteristic strain spectrum} \cite{Maggiore} or {\it chirp amplitude} \cite{Boyle}.
The lower and upper limit of integration in Eq.(\ref{gwe})
can be taken to be  $f_{low}\simeq f_E$
and  $f_{upper}\simeq f_1$, respectively, since only the wavelength of the modes inside the horizon contribute to the total energy density.

In the present universe,
the physical frequency relates to a comoving wave number  $k$ as
\be \label{freq}
f=  \frac{k}{2\pi a (\tau_0)} = \frac{k}{2\pi l_H}.
\ee
 Thus, from Eqs. (\ref{frelation}) and  (\ref{freq}) one can easily get the each characteristic  physical frequency: $f_H=H_0\simeq 2.28\cdot10^{-18}$ Hz,
 $f_E\simeq1.93\cdot10^{-18}$ Hz,
 $f_2\simeq 9.3\cdot10^{-17}$ Hz. Moreover,
 $f_s\simeq 2.14\cdot10^8$ Hz  and $f_1\simeq4.71\cdot10^8$ Hz for $n_s=0.967$;
 $f_s\simeq 1.03\cdot10^6$ Hz  and $f_1\simeq8.04\cdot10^7$ Hz for $n_s=0.966$.
  The values of $f_1$ are below the  constraint  from the rate of
 the primordial nucleosynthesis, $f_1\lesssim3\times10^{10}$ Hz
 \cite{grishchuk}. When the acceleration epoch is considered, the constraint becomes
  $f_1\simeq4\times10^{10}$ Hz. Note that, Eq. (\ref{frelation}) implies that $f_1$ depends on
 the value of $\beta_s$ but not $\beta$. Therefore, $f_1$
 is a fixed value once   $\beta_s$ has been chosen.

\section{ Observational constraints  and the detection}

As well known, the anisotropies and polarizations of CMB are contributed by two parts: tensor metric perturbations (gravitational waves)
and curvature perturbations. Moreover, the B-mode CMB polarization is only produced by tensor perturbations.
 In literatures \cite{Peiris,Spergel,Hinshaw,Komatsu}, the tensor-to-scalar ratio $r$ is often
  introduced
\be\label{ratio}
r=\frac{\Delta^2_h(k_0)}{\Delta^2_{\mathcal{R}}(k_0)},
\ee
where $\Delta^2_h(k_0)$ and $\Delta^2_{\mathcal{R}}(k_0)$ are the power spectrum of the tensor perturbations and
curvature perturbations evaluated at the pivot wavenumber $k^p_0=k_0/a(\tau_0)=0.002$ Mpc$^{-1}$ \cite{Komatsu}, respectively. The corresponding physical frequency is
$f_0\simeq3.09\times10^{-18}$ Hz. For tensor power spectrum, one has the definition:
\be
\Delta^2_h(k)\equiv h^2(k,\tau_0).
\ee
Hence, the non-zero value of $r$ implies the existence of gravitational wave background, and would be probed
 with measurements of the B-mode CMB polarization \cite{Amarie}. Even though there is still no direct observation of $\Delta^2_h(k_0)$, $\Delta^2_{\mathcal{R}}(k_0)$ can be fixed by CMB observations,
$\Delta^2_{\mathcal{R}}(k_0)\equiv A_s=(2.43\pm 0.11)\cdot10^{-9}$
by WMAP\,7 Mean \cite{Komatsu}. On the other hand, at present only
observational constraints on $r$ have been given by WMAP
\cite{Komatsu,Hinshaw}. The upper bounds of $r$ are
recently constrained \cite{Komatsu}  as  $r<0.24$ by WMAP+BAO+$H_0$ and $r<0.36$ by
WMAP 7 only for the vanishing scalar running spectral index $\alpha_s$, and  $r<0.49$  for the  non-vanishing $\alpha_s$
by both the combination of WMAP+BAO+$H_0$ and the WMAP 7 only,
respectively. Furthermore, using a discrete, model-independent measure of the degree
of fine-tuning required, if $0.95\lesssim n_s<0.98$, in accord with current measurements,
 the tensor-to-ratio satisfies $r\gtrsim10^{-2}$  \cite{Boyle2}.
Therefore, one can normalize the RGWs at $k=k_0$ using Eq. (\ref{ratio}), if $r$ can be determined definitely.


Since $k_H\leq k_0\leq k_2$,
  $h(k_0,\tau_0)$ re-entered the horizon during the matter domination and  is now inside  the present horizon.
   Therefore, $h(k_0,\tau_0)$ suffered  a decay $\propto 1/a$ from   the horizon re-entry to the present time.
According to  Eq. (\ref{htauh}) one has
\be\label{primordial}
h(k_0,\tau_0)\equiv\Delta_h(k_0) =A\left(\frac{k_0}{k_H}\right)^{\beta}(1+z_E)^{-\frac{2+\gamma}{\gamma}}.
\ee
 Therefore, $A$ can be fixed for a determined value of $\Delta_h(k_0)$, given a definite value of $\beta$.
Note that,   the ``thin-horizon'' approximation that treats
horizon re-entry as a ``sudden'' or instantaneous event \cite{Boyle} has been used in Eq.(\ref{primordial}).
Thus, Eq.(\ref{primordial}) is a normalization of RGWs provided by the observations of  CMB. On the other hand,
from the point of view of theories, the initial condition of RGWs could also be given, or some indirect connections
between different parameters would exist. In the following, we discuss two different theories.

\begin{figure}
\resizebox{180mm}{!}{\includegraphics{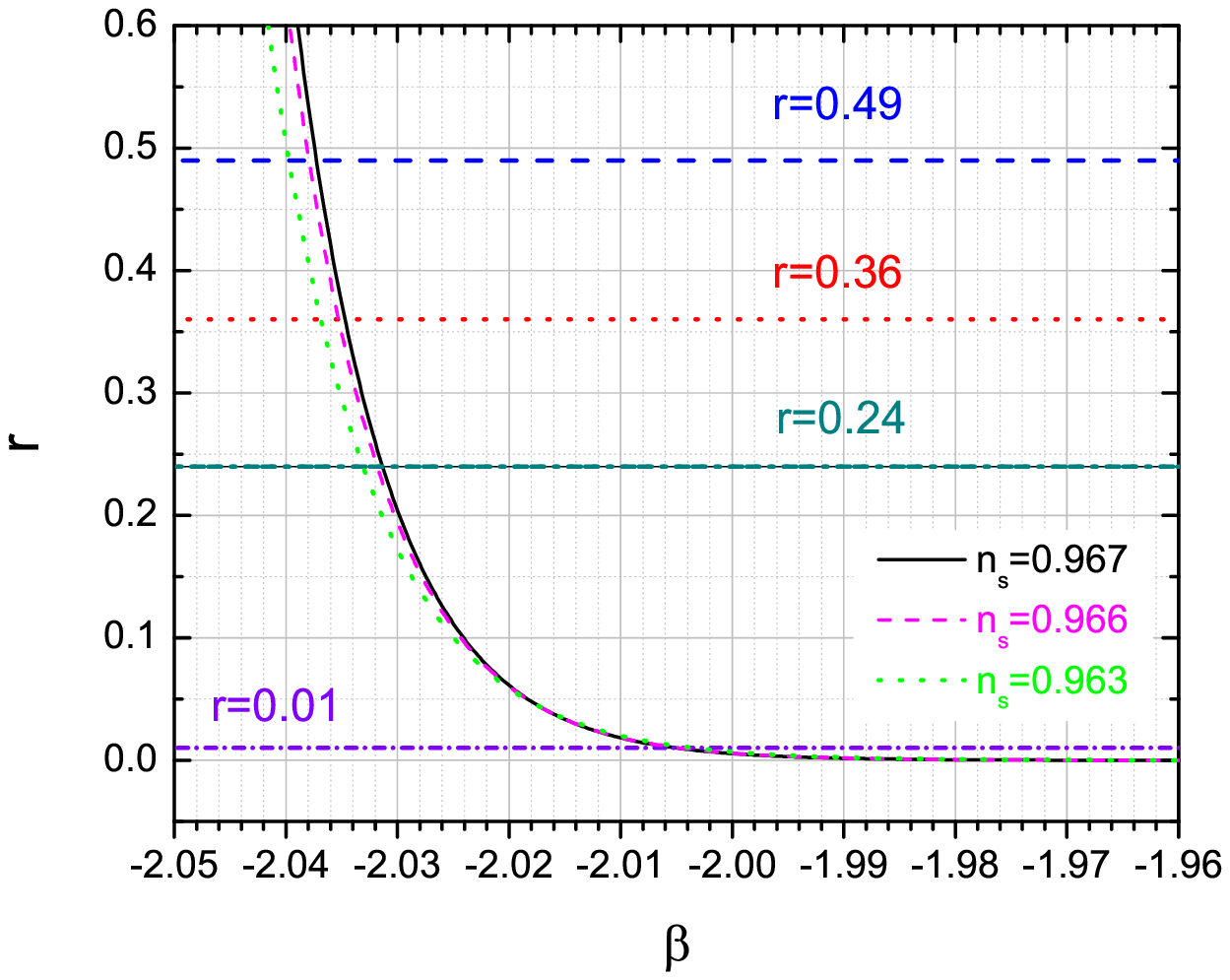}
\includegraphics{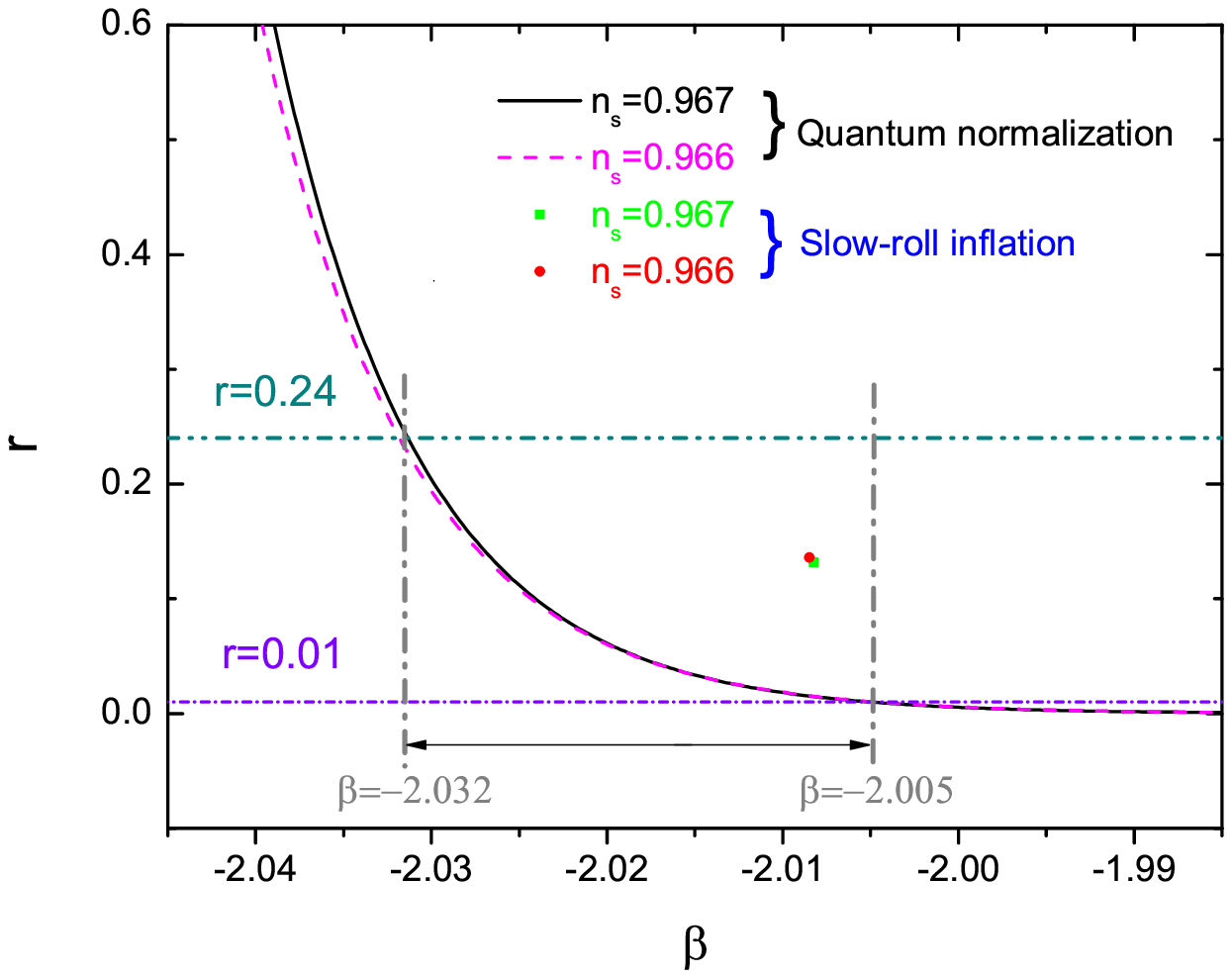}}
\caption{\label{rbeta}
The relation between $r$ and $\beta$. Left: Three $r-\beta$ curves based on the quantum normalization for
three different values of the scalar spectral index $n_s=0.967$,  $n_s=0.966$ and $n_s=0.963$, respectively.
Right: The comparison between the $r-\beta$ relations based on the quantum normalization and those based on the slow-roll inflation.
}
\end{figure}

Firstly, the constant $A$ appearing in Eq. (\ref{initialspectrum})  can be  determined by quantum normalization \cite{grishchuk}:
\be\label{aini}
 A=b8\sqrt{\pi}\frac{l_{Pl}}{l_0},
  \ee
  where $b\equiv\gamma^{2+\beta}/|1+\beta|^{1+\beta}$  in our notation,  $l_{Pl}=\sqrt{G}$ is the
 Plank length, and $l_0$ has been fixed  to be \cite{Miao}
  \be\label{l0}
  l_0=b H_0^{-1} \zeta_1^{\frac{\beta-\beta_s}{1+\beta_s}}\zeta_s^\beta \zeta_2^{\frac{\beta-1}{2}}
  \zeta_E^{-(1+\frac{1+\beta}{\gamma})},
   \ee
   using the continuous conditions of $a(\tau)$ and $a'(\tau)$ at the joining moments between two
   different phases. Here, it should be  point out that the values of $\zeta_s$ and $\zeta_1$ shown in
   Eq. (\ref{l0})  will be chosen tentatively as those determined by Eqs. (\ref{zetas}) and (\ref{delta2}), respectively,
   since they are not known well.
With the help of
  Eqs. (\ref{ratio}), (\ref{aini})   and (\ref{l0}), Eq. (\ref{primordial}) reduces to
\be\label{arelation}
\Delta_\mathcal{R}(k_0)r^{1/2}=8\sqrt{\pi}l_{Pl}H_0
\zeta_1^{\frac{\beta_s-\beta}{1+\beta_s}}\zeta_s^{-\beta} \zeta_2^{\frac{1-\beta}{2}}
  \zeta_E^{\frac{\beta-1}{\gamma}}\left(\frac{k_0}{k_H}\right)^\beta,
\ee
which implies a relationship between $\beta$, $\beta_s$ and $r$.
Given $\beta_s=1$, we illustrate  $r$ as a function of $\beta$ in the left panel of Fig. \ref{rbeta} for $n_s=0.967$,
$0.966$ and $0.963$, respectively. Here we plus the case of $n_s=0.963$ for comparison. One
  can see that the  $r-\beta$ curves are almost overlap for $\beta\gtrsim-2$ and
   only small discrepancies exist for   lower  $\beta$. Taking  $n_s=0.966$ for example,  the constraints  $r<0.49$,  $r<0.36$ and $r<0.24$ lead to  $\beta\gtrsim-2.038$,  $\beta\gtrsim-2.035$, and    $\beta\gtrsim-2.032$, respectively, while $r>0.01$ gives $\beta\lesssim-2.005$.
Note that, these results are based on the validity  of quantum normalization Eq. (\ref{aini}), however, it is
not the unique initial condition. This is a manifestation of the  vacuum ambiguity
that is responsible for particle production in cosmological spacetimes \cite{Birrell}. Furthermore, we have
taken the values of $\zeta_s$ and $\zeta_1$ evaluated by using the slow-roll inflation model with a scalar potential
$V(\phi)=\frac{1}{2}m^2\phi^2$ \cite{Mielczarek}. If  other different values of $\zeta_s$ and $\zeta_1$ are chosen,
we find that  the curves plotted in Fig. \ref{rbeta} will change sensitively. For simplicity, in this paper we will not consider
  these cases.

 Secondly, from the point of view of the slow-roll inflation model, there is a natural relation between $n_s$ and $r$.
 For $V(\phi)=\frac{1}{2}m^2\phi^2$, using Eq. (\ref{slow}) one easily has $\epsilon=\eta$ and
 \be\label{ns}
 n_s\approx1-4\epsilon.
 \ee
On the other hand, under the slow-roll approximation, the primordial tensor power spectrum and the primordial
scalar power spectrum are respectively given as \cite{Boyle,Kuroyanagi}:
\ba\label{tensor}
&&\Delta_h^2(k,\tau_\ast)\approx\frac{16}{\pi}\left(\frac{H_\ast}{m_{\rm{Pl}}}\right)^2,\\
&&\Delta_\mathcal{R}^2(k,\tau_\ast)\approx\frac{1}{\pi\epsilon}\left(\frac{H_\ast}{m_{\rm{Pl}}}\right)^2,\label{scalar}
\ea
where $H_\ast$ is the   Hubble rate during the inflation stage and is invariable  for the slow-roll approximation.
$\tau_\ast$ stands for the moment when the $k$-mode exits the horizon.
According to the original definition of $r$, i.e., the ratio of the primordial tensor power spectrum to the primordial scalar power spectrum, one has \cite{Boyle,Kuroyanagi}
\be\label{ratio2}
r\equiv \Delta_h^2(k,\tau_\ast)/\Delta_\mathcal{R}^2(k,\tau_\ast)=16\epsilon,
\ee
with the help of Eqs. (\ref{tensor}) and (\ref{scalar}). Note that the definition of $r$ in Eq.(\ref{ratio2}) is
little different from that in Eq. (\ref{ratio}). We will discuss the difference below. In the literatures of WMAP \cite{Peiris,Spergel,Komatsu}, the primordial
power spectrum are often written as
\ba \label{tensor2}
&&\Delta_h^2(k)=\Delta_h^2(k_0)\left(\frac{k}{k_0}\right)^{n_t},\\
&&\Delta_\mathcal{R}^2(k)=\Delta_\mathcal{R}^2(k_0)\left(\frac{k}{k_0}\right)^{n_s-1}\label{scalar2},
\ea
where  $n_t$ is the tensorial spectral index. Generally, $n_t$
is   $k-$depedent \cite{grishchuk91,Kosowsky,LiddleA,TongZhang},
however, for simplicity we just consider $n_t$ as a constant in this paper. With the
help of Eq. (\ref{initialspectrum}), one easily find that $n_t=2\beta+4$.
Strictly speaking, based on Eq. (\ref{htaue}), Eq. (\ref{tensor2}) is valid approximately since the mode $k=k_0$ has
re-entered the horizon at the present time. An analogous case would also exist in  Eq. (\ref{scalar2}), which
 was used to evaluate the reheating temperature in \cite{Mielczarek}.
 However, as a conservative consideration, Eqs. (\ref{tensor2}) and (\ref{scalar2}) provide a good approximation to evaluate $r$. Now let us estimate the discrepancy between Eq. (\ref{ratio}) and (\ref{ratio2}) in the following. From Eq. (\ref{tensor2}),
it is straight forward to get $r\equiv\frac{\Delta_h^2(k_E)}{\Delta_\mathcal{R}^2(k_E)}=\frac{\Delta_h^2(k_0)}
{\Delta_\mathcal{R}^2(k_0)}(k_E/k_0)^{n_t+1-n_s}$. The differential factor $(k_E/k_0)^{n_t+1-n_s}\sim0.992$
for $n_s=0.966$ or $n_s=0.967$, where we have used Eq. (\ref{ns}) and the well known relation
\be\label{nt}
n_t\approx-2\epsilon
\ee
in the slow-roll inflation.
Allowing for the relation $n_t=2\beta+4$, the combination of  Eqs. (\ref{ratio2}) and (\ref{nt}) give
\be\label{rbeta3}
r=-8n_t=-16(\beta+2).
\ee
This general linear relation is plotted in the right panel of  Fig. \ref{rbeta}. For a fixed value of $n_s$,    $r$ can be determined
 through Eqs. (\ref{ns}) and (\ref{ratio2}), and then $\beta$ can also be determined through Eq. (\ref{rbeta3}) correspondingly. We employ a red circle   and a green square   to stand
 for the values of $(\beta,r)=(-2.009,0.136)$ and  $(\beta,r)=(-2.008,0.132)$ corresponding to the observations $n_s=0.966$ and $n_s=0.967$, respectively. For comparison, the relation between $r$ and $\beta$ determined by the
 quantum normalization was shown again, and the constraint  of $0.01<r<0.24$ discussed above leads to $\beta$ localizing
 in the range of   $-2.032<\beta<-2.005$. On the other hand, according to Eq.  (\ref{rbeta3}), the constraint  of $0.01<r<0.24$
 forces $\beta$ to lie in the range of $-2.015<\beta<-2.001$.  Hence, under the condition $r\gtrsim0.01$, $\beta$ has a tilt  smaller than $-2$ both for the two   theoretical bases.    In fact, Eq. (\ref{rbeta3}) directly implies $\beta<-2$ due to the definition of $r$ being positive.

\begin{figure}
\resizebox{180mm}{!}{\includegraphics{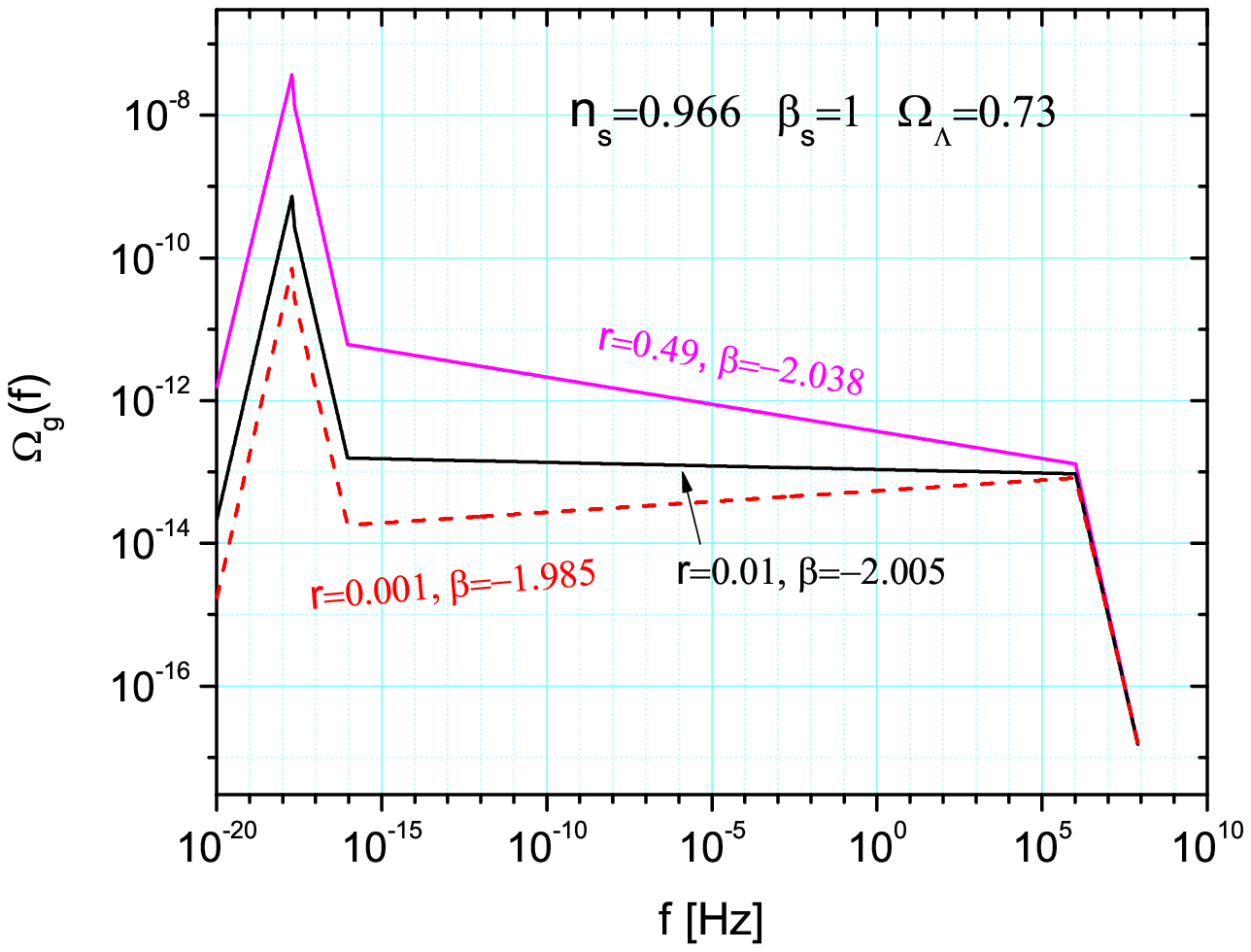}
\includegraphics{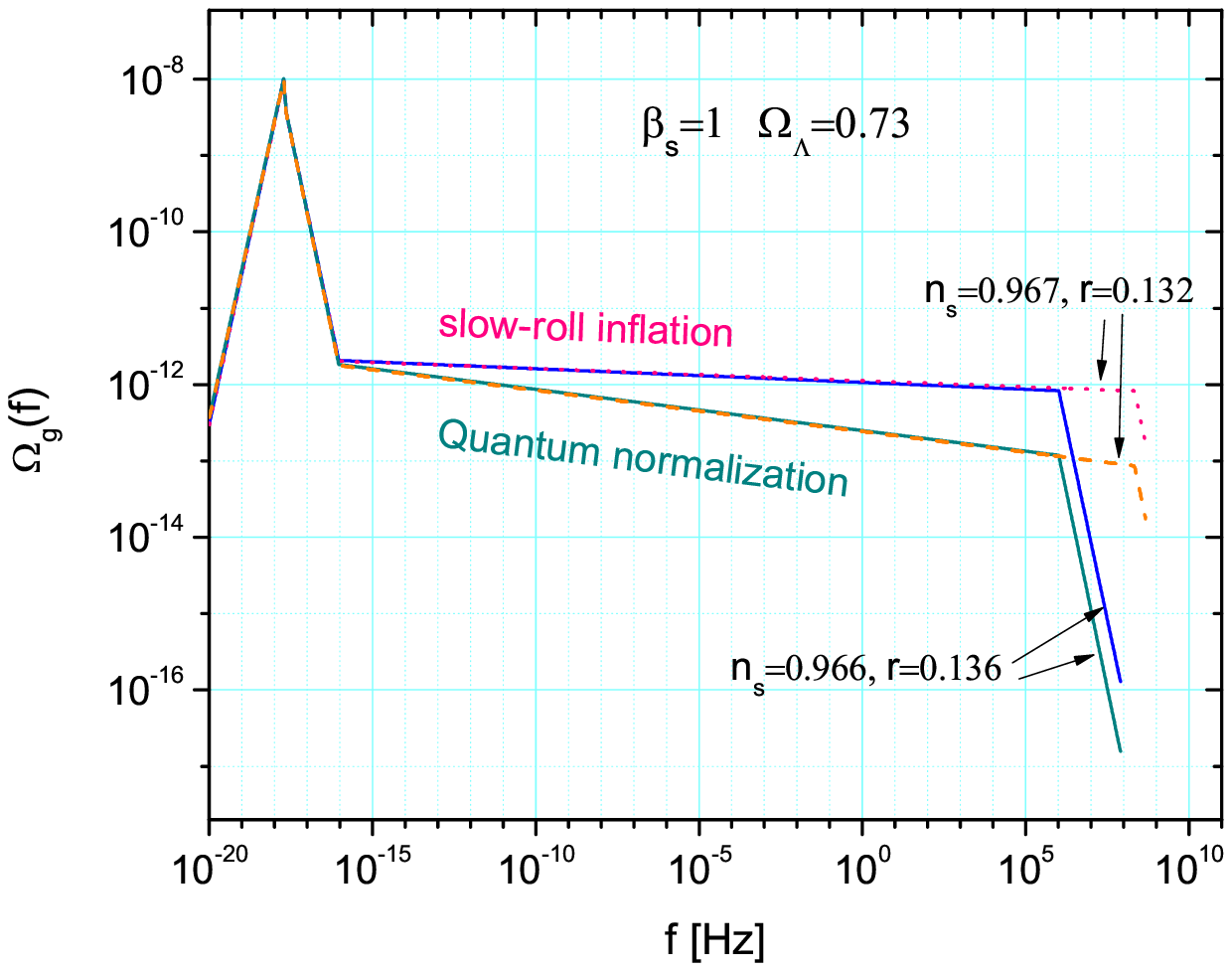}}
\caption{\label{relic}
Left: The energy density  spectra of RGWs based on the quantum normalization with $r=0.49(\beta=-2.038)$, $r=0.01(\beta=-2.005)$ and $r=0.001(\beta=-1.985)$, respectively, for $n_s=0.966$.
 Right: The comparisons between the energy density spectra based on  the quantum normalization and
 the slow-roll inflation for $n_s=0.966$ and $n_s=0.967$, respectively. }
\end{figure}

In the following, we demonstrate the properties of  the energy spectra of RGWs taking some definite values of $r$ and $\beta$ for
 example.  Firstly, setting $n_s=0.966$, the approximate energy spectra with
three different tensor-to-scalar ratios $r=0.49$, $r=0.01$ and $r=0.001$, which correspond to $\beta=-2.038$, $\beta=-2.005$ and $\beta=-1.985$ according to  quantum normalization,  respectively, are illustrated in the left panel of Fig. \ref{relic}.
It can be found that all the three cases  of energy spectrum have the same amplitudes
 at $f= f_1\simeq8.04\cdot10^7$ Hz. This is a natural result from the quantum normalization.
Plugging Eq. (\ref{aini}) into Eq. (\ref{hf1})  with the help of  Eqs. (\ref{frelation}) and (\ref{l0}), one obtain
\be
h_c(f_1)=4\sqrt{2\pi} f_1 l_{Pl},
\ee
and it is straight forward to obtain
\be
\Omega_g(f_1)=\frac{64\pi^3f_1^4l_{Pl}^2}{3H_0^2},
\ee
according to Eq. (\ref{omega}).
 We can see that $h_c(f_1)$ and $\Omega_g(f_1)$ are fixed  values independent on $\beta$, since $f_1$ is independent on $\beta$. On the other hand, as shown in Fig. \ref{relic}, a smaller  $\beta$ leads to larger energy spectrum   especially at lower frequencies.
 This property is contrary to that illustrated in Refs. \cite{Miao,TongZhang,zhangtong},
where a fixed $\beta_s$ was chosen but the quantum normalization was not employed.
Secondly, let us see the differences  between the energy spectra of RGWs based on
 the quantum normalization and those based on the slow-roll inflation. In the right
 panel of Fig. \ref{relic}, we plot the energy spectra of RGWs based on the
 two different theories  for $n_s=0.966$ and $n_s=0.967$, respectively.
For the slow-roll inflation, once $n_s$ is given, $r$ and $\beta$ will be determined
subsequently. Concretely, as analyzed above, $n_s=0.966$ and $n_s=0.967$ correspond to
$r=0.136$ and $r=0.132$, respectively. For comparison, we also choose the same values of  $r$  in the
energy spectra   based on   quantum normalization, and moreover,  it is easy to have $\beta\simeq-2.027$ for $r=0.136$ or $r=0.132$ using Eq. (\ref{arelation}). As shown in the right panel of Fig. \ref{relic}, the energy spectra of RGWs
based on these  two theories are almost the same at $f\leq f_2$ both for $n_s=0.966$ and $n_s=0.967$. This is
because of the normalization from the CMB observations at $f=f_0$, where  $\Omega_g(f)/r $ is a fixed value.
Therefore, around $f=f_0$, $\Omega_g(f)$ depends on $r$ sensitively rather than $\beta$, since $\Omega_g(f)$ depends on $\beta$ in the way of $\Omega(f)\propto(f/f_0)^{2\beta}$. Let us take $\Omega_g(f_2)$ for example.
With the help of Eqs. (\ref{htauh}), (\ref{omega}) and (\ref{primordial}), one can easily derive that
$\Omega_g(f_2)\propto r(\frac{f_2}{f_0})^{2\beta}(\frac{f_2}{H_0})^2$. Among the four
curves illustrated in the right panel of Fig.\ref{relic},  the biggest relative discrepancy of $\Omega_g(f_2)$, existing
 between the sets $(r=0.136,\beta=-2.008)$ and $(r=0.132,\beta=-2.027)$,   is about  $17\%$.
Similarly, for the modes which re-entered the horizon during radiation domination, one has $\Omega_g(f)\propto r(\frac{f}{f_0})^{2\beta}(\frac{f}{H_0})^4$ due to Eq. (\ref{hpulsar}). Therefore, for fixed $r$, the discrepancies of the energy spectra based on quantum normalization and those based on slow-roll inflation are larger
 and larger with the increasing frequency, since a difference $\Delta\beta\simeq0.02$ in the indices between them always exists.
 In the very high frequency band $f_s<f<f_1$, not only  the discrepancy discussed  above    is larger,
 but also  the energy spectra manifest
 different properties for different values of $n_s$. The latter phenomenon results from that   $n_s=0.966$ and  $n_s=0.967$    lead  to different values of $f_s$ and $f_1$ as analyzed below Eq. (\ref{freq}), either on the basis of
quantum normalization or slow-roll inflation. The value of $f_s$ depends on
 $n_s$ very sensitively, while $f_1$ depends on $n_s$ much less sensitively compared to $f_s$. These properties would become
 an effective tool to discriminate the value of $n_s$, and could be an interesting object of the high-frequency
 gravitational wave detectors \cite{cruise,fangyu,Akutsu}.

 \begin{figure}
\resizebox{100mm}{!}{\includegraphics{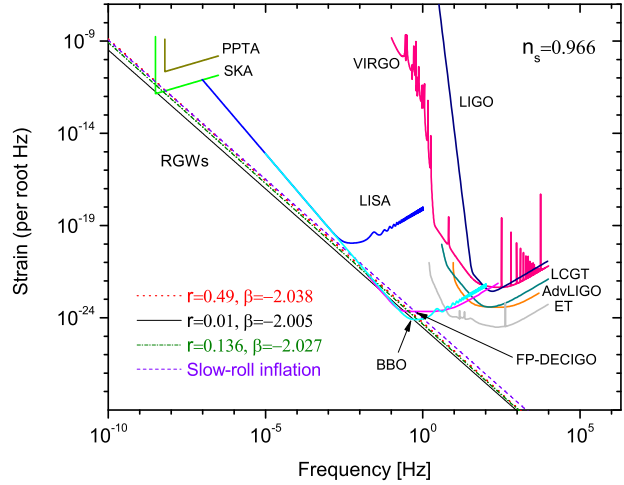}}
\caption{\label{detection}
The strain of RGWs with different parameters for $n_s=0.966$ confronts against the current and planed GW detectors. The sensitivity curves of PPTA and SKA using pulsar timing technique are taken from Refs.\cite{Jenet} and \cite{Sesana}, respectively.
The curve of BBO is generated using the online ``Sensitivity curve generator'' \cite{lisa} with the parameters in Table II of Ref.\cite{Crowder} and Table I of Ref.\cite{Cutler}. The curve  of FP-DECIGO is taken from Ref.\cite{Kawamura}. The sensitivity curves of the ground-based interferometers AdvLIGO, LCGT and ET are taken from Refs.\cite{advligo}, \cite{lcgt} and \cite{Hild}, respectively. }
\end{figure}
\begin{figure}
\resizebox{180mm}{!}{\includegraphics{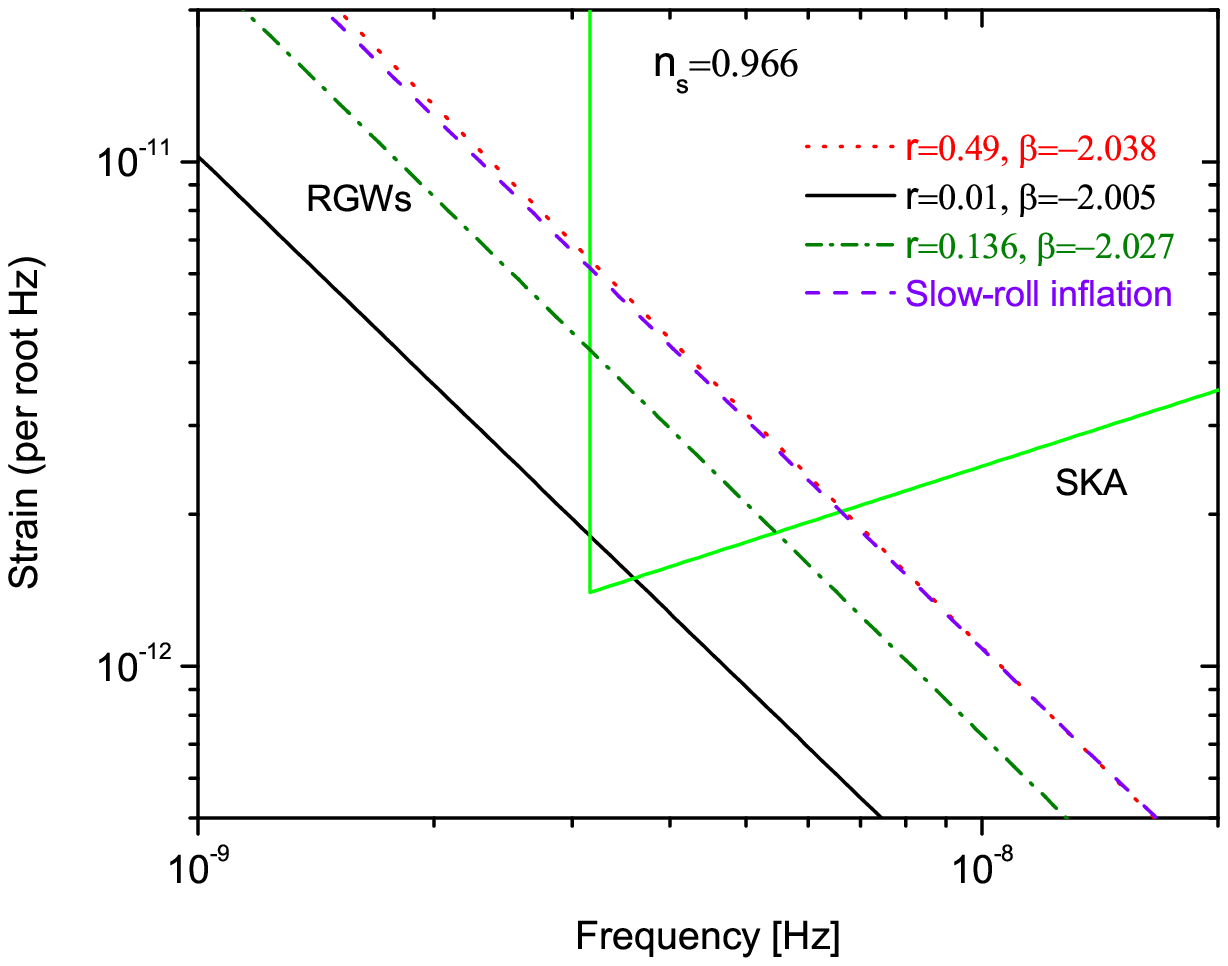}
\includegraphics{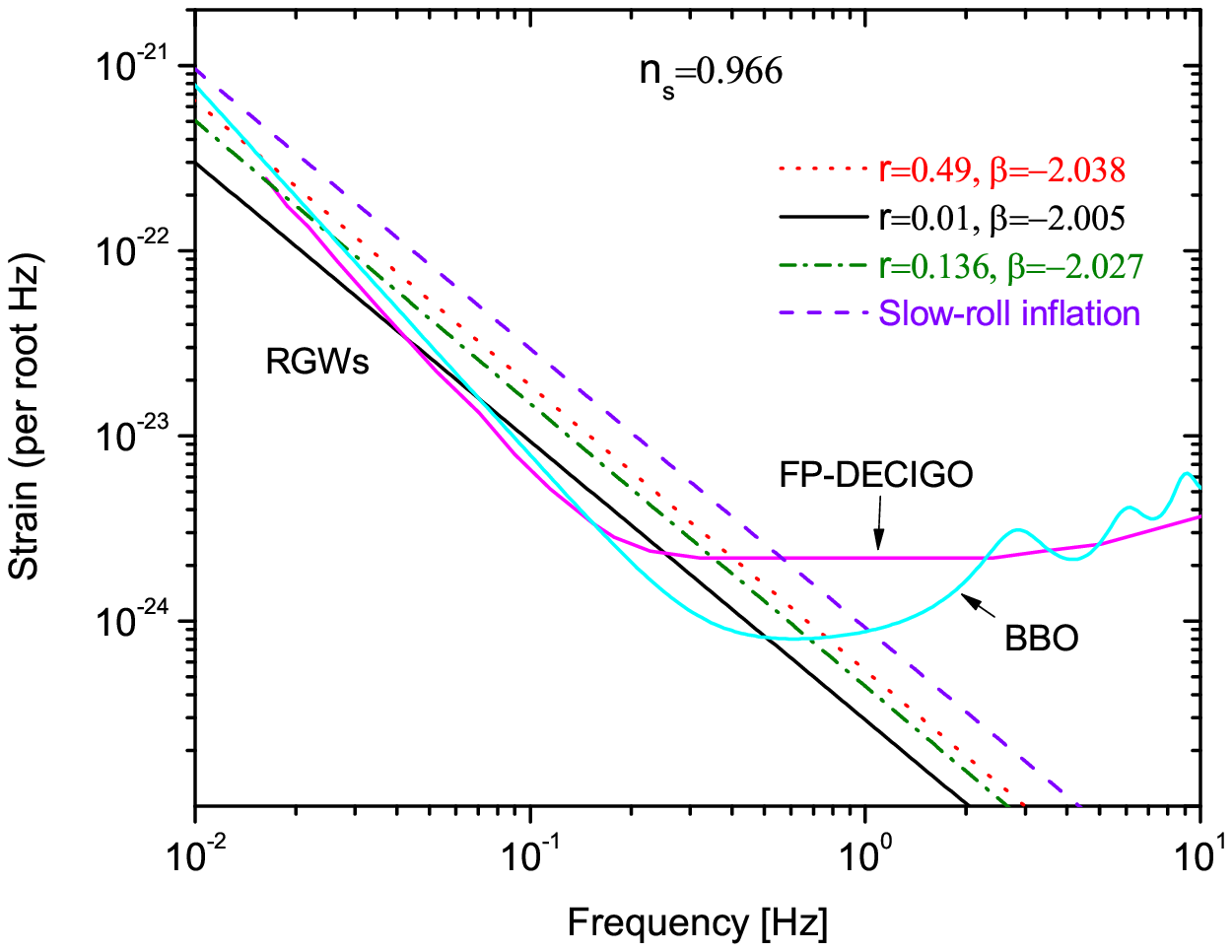}}
\caption{\label{detection2}
A zoom plot of the two parts of SKA¡¡ and  FP-DCIGO/BBO. All the values of parameters in RGW models and the sensitivity curves of the three gravitational wave  detectors are exactly the same as those shown in Fig. \ref{detection}. }
\end{figure}

Below, let us discuss the detection of RGWs using the ongoing and planed gravitational detectors which are sensitive at different  frequency bands. As shown in the right panel of Fig. \ref{relic}, there is nearly no difference  between the RGWs   with $n_s=0.966$ and those with  $n_s=0.967$ at $f<f_s$, both based on quantum normalization and slow-roll inflation. Thus,
we just take $n_s=0.966$ for demonstration in the following.
As a conservative evaluation, in Fig. \ref{detection} we show  the strain sensitivity curves  of
 various gravitational wave detectors and the strain amplitudes, $h_c(f)/\sqrt{f}$,
  of the RGWs for $n_s=0.966$ based on quantum normalization and slow-roll inflation, respectively.
These detectors contain  the PPTA \cite{Jenet} and SKA \cite{Sesana} using the pulsar timing technique,  the
 space-based laser interferometers such as LISA \cite{lisa},  BBO \cite{Crowder,Cutler}, and the Fabry-Perot DECIGO \cite{Kawamura},  the  ground-based laser interferometers
 including the first
 generation   LIGO \cite{ligo1} and VIRGO \cite{virgocurve}, the second generation  AdvLIGO \cite{advligo} and LCGT \cite{lcgt}, and the third generation ET \cite{Hild}. One can see that
 all the ground-based interferometers can hardly detect the theoretical RGWs discussed above. PPTA and LISA also
 have difficulties to catch RGWs, however, the planed
 SKA, BBO and DECIGO are promising to detect RGWs since they have relative higher sensitivities.
  In order to show the detectabilities of RGWs by SKA and DECIGO/BBO more clearly, we enlarge the
 two parts of SKA and DECIGO/BBO which are shown in Fig. \ref{detection2}.
 As can be seen in the left panel of Fig. \ref{detection2}, around the frequencies $3-5\times10^{-9}$ Hz,  the spectra of
 RGWs with different parameters of $r$ (and in turn the different corresponding $\beta$ due to the $r-\beta$ relation) based on quantum normalization have
 different amplitudes, and they are promising to be distinguished by SKA.
 However, the spectrum based on
 slow-roll inflation is hardly distinguished from that based on quantum normalization with $r=0.49(\beta=-2.038)$
 by SKA, since two lines   almost overlap in the frequency response band of SKA. On the other hand, in the right panel of Fig. \ref{detection2} one can see that,
    RGWs based on slow-roll inflation are easier to be distinguished
 from those based on quantum normalization by DECIGO or BBO in the frequency range  $\sim10^{-2}-10^0$ Hz. Furthermore, since DECIGO/BBO has a wider frequency response band in contrast to SKA, the $r-\beta$ has the potential to be examined. In conclusion, the combination of SKA and DECIGO/BBO
 provides an important tool not only distinguishing different theoretical models of RGWs but also determining
 the corresponding  parameters.

\section{Discussions}

~We determined
the expansion histories of the preheating stage and
the radiation-dominated stage due to the  the reheating temperature which is recalculated using the  latest observations of WAMP 7-year. Based on that, we illustrated the approximate solutions of RGWs in the current accelerating stage for
various frequency bands. We found that  the frequency $f_s$, describing that the mode re-entered the horizon at the moment of the reheating,
is  dependent on $n_s$ sensitively, however, the upper limit frequency of RGWs depends  on the value of
$n_s$ much less sensitively.
  Combing the quantum normalization of RGWs with
the CMB observations, we obtained a relation between the tensor-to-scalar ratio $r$ and the
inflation index $\beta$ for the fixed   preheating index $\beta_s=1$. According to the relation between
$r$ and $\beta$ with a fixed $n_s=0.966$, we find that a relatively tight constraint  $0.01<r<0.24$  leads to $\beta$ localizing
 in a range of   $-2.032<\beta<-2.005$ based on quantum normalization and
 in a more narrow range of  $-2.015<\beta<-2.001$ based on slow-roll inflation, respectively.
  We  plotted the spectrum and the energy density spectrum of the RGWs for three cases of $r=0.49(\beta=-2.038)$,
$r=0.01(\beta=-2.005)$, and $r=0.001(\beta=-1.985)$, respectively. It was found that a lager $r$, i.e., a smaller $\beta$ leads to a larger spectrum of RGWs
especially at lower frequencies. For comparison, we also illustrated the spectra of RGWs with the parameters of
 $r$ and $\beta$ given by the slow-roll inflation. Concretely, for $n_s=0.966$, one has $r=0.136$ and $\beta=-2.009$,
 and for $n_s=0.967$, one has $r=0.132$ and $\beta=-2.008$.
 It was found that, for the same value of $r$, the discrepancy of the energy spectra based on quantum normalization and those based on slow-roll inflation is larger
 and larger with the increasing frequency. However, our analysis above does not not apply, in general, to less conventional models of inflation where the RGW spectrum and the observed spectrum of scalar  perturbations are produced with different primordial mechanisms \cite{Bozza,Gasperini}, and, as a consequence, they are in principle completely decoupled.

 Among the current and planed GW detectors, only  the planed SKA using the pulsar timing technique, and the planed space-based interferometers BBO and  DECIGO    are promising to detect RGWs. However, these results are based on
 the referenced values of $\zeta_1$ and $\zeta_s$ which are obtained from the combination of CMB observations and the slow-roll inflation with a concrete potential $V(\phi)=
 \frac{1}{2}m^2\phi^2$.
Note that, the values of  $\zeta_1$ and $\zeta_s$ depend   sensitively on the reheating temperature $T_{\rm{RH}}$
 which is dependent on $n_s$ sensitively. However, the $r-\beta$ relation is nearly not dependent on $n_s$ once
 the form of the potential $V(\phi)= \frac{1}{2}m^2\phi^2$ is given. Hence, for the same value of $r$, different
 $n_s$ only affects the very high frequency  RGWs that re-entered the horizon before the radiation-dominant  stage.
Therefore, to a certain extant, our analysis of the detection in the frequency range $f\leq10^3$ Hz is general,   even though the  determination of the reheating temperature  from CMB has very large uncertainties.  As forecasted in Ref. \cite{Mielczarek}, the future CMB experiments such as the Plank satellite \cite{Planck}, the ground-based  ACTPol  \cite{Niemack} and the planned CMBPol \cite{CMBpol} will provide  significant reductions of the uncertainties of the reheating temperature $T_{\rm{RH}}$. Therefore, one expect accordingly that the expansion histories of the very early universe would be known better.
 On the other hand,  the values of $\zeta_1$ and $\zeta_s$ could be chosen independently on the slow-roll inflation scenario,    which would be studied in the future work.

\

{ACKNOWLEDGMENT}: This work is supported by the National Science Foundation of China  under Grant No. 11103024,
and has been  supported in part by  the Fundamental Research Fund of Korea Astronomy and Space Science Institute.


\begin{thebibliography}{88}
\bibitem{here} Here, the superscript
$p$ means``physical'', and we use $k$ standing for the conmoving number.
\end{thebibliography}
\end{document}